\documentclass[11pt]{article}
\usepackage[margin=1in]{geometry}
\usepackage[T1]{fontenc}
\usepackage{lmodern,microtype}
\usepackage{amsmath,amssymb,amsthm,mathtools}
\usepackage[dvipsnames]{xcolor}
\usepackage{tikz}
\usetikzlibrary{arrows.meta,decorations.pathreplacing}
\usepackage{enumitem,needspace}
\usepackage[colorlinks=true,linkcolor=blue!60!black,citecolor=green!40!black,urlcolor=blue!60!black]{hyperref}
\usepackage[capitalise,nameinlink]{cleveref}
\hypersetup{pdftitle={Amortized Cell-Probe Lower Bounds for Dynamic Boolean Problems},pdfauthor={Young Kun Ko}}

\crefname{Protocol}{Protocol}{Protocols}

\theoremstyle{plain}
\newtheorem{theorem}{Theorem}[section]
\newtheorem{lemma}[theorem]{Lemma}
\newtheorem{claim}[theorem]{Claim}
\newtheorem{fact}[theorem]{Fact}
\newtheorem{corollary}[theorem]{Corollary}
\crefname{corollary}{Corollary}{Corollaries}
\crefname{fact}{Fact}{Facts}
\crefname{claim}{Claim}{Claims}

\newcommand{\eps}{\varepsilon}
\newcommand{\E}{\mathop{\mathbb{E}}}
\newcommand{\abs}[1]{\left|#1\right|}
\providecommand{\norm}[1]{\left\lVert #1\right\rVert}

\newcommand{\cI}{\mathcal{I}}
\newcommand{\cE}{\mathcal{E}}

\newcommand{\DISJ}{\mathrm{DISJ}}
\newcommand{\FAIL}{\mathrm{FAIL}}
\newcommand{\vS}{\vec{S}}
\newcommand{\Csam}{\widetilde{C}_i}
\newcommand{\Aall}{A_{\mathrm{all}}}
\newcommand{\polylog}{\operatorname{polylog}}
\newcommand{\Patrascu}{P\u{a}tra\c{s}cu}

\title{Amortized Cell-Probe Lower Bounds \\ for Dynamic Boolean Problems}
\author{Young Kun Ko\thanks{Department of Computer Science and Engineering, Pennsylvania State University, University Park, PA, USA. \texttt{ykko@psu.edu}}}
\date{November 5, 2026}

\begin{document}
\maketitle

\begin{abstract}
We prove the first unconditional amortized $\Omega\big((\log N/\log\log N)^2\big)$ cell-probe lower bound for a Boolean dynamic problem. For insert-only multiphase Disjointness with $N$ query sets over $[N]$ and word size $w=\Theta(\log N)$, a fixed distribution of interleaved insertions and queries forces this expected average cost per operation, excluding preprocessing. The bound allows sufficiently small constant error for each query of every fixed legal sequence and imposes no restriction on individual operation costs. It transfers to incremental directed reachability, incremental subgraph connectivity, and fully dynamic triangle detection.

We also prove $t_q=\Omega\big((\log N/\log(t_uw))^2\big)$ for amortized insertion time $1\le t_u=N^{o(1)}$ and expected query time $t_q$, when the number of insertions before the final query is not announced in advance. We reach the quadratic-logarithmic scale with bounded error by combining interleaving with the 2.5-round communication framework of \cite{ko_omega_2026,ko_omegalog_2026} and controlling errors among accepted simulations.
\end{abstract}

\newpage
% =====================================================================
\section{Introduction}\label{sec:intro}
% =====================================================================

We prove the first unconditional amortized $\Omega\big((\log n/\log\log n)^2\big)$ cell-probe lower bound for a Boolean dynamic problem, whose queries return a single bit. The bound holds for randomized data structures with sufficiently small constant error on each query of every fixed legal operation sequence. It charges the total work of interleaved updates and queries, with no worst-case bound on individual operation costs.

Super-logarithmic amortized Boolean lower bounds can already be obtained with zero error by adapting the method of Larsen, Weinstein and Yu \cite{larsen_crossing_2020}. For Boolean polynomial evaluation, interleaving and averaging over a random query time give an $\Omega(\log^{3/2}n/(\log\log n)^2)$ amortized bound; we explain this consequence of their method in \cref{sec:previous}. Our contribution is to reach the quadratic-logarithmic scale and allow constant per-query error.

We prove the bound for insert-only multiphase Disjointness. In the multiphase problem of \Patrascu~\cite{patrascu_mihai_towards_2010}, a data structure preprocesses $m$ sets $S_1,\dots,S_m$, receives updates to a set $X$, and answers whether $S_q\cap X=\emptyset$ for a given $q\in[m]$. For Disjointness, \cite{ko_omegalog_2026} proved the trade-off $t_q=\Omega\big((\log n/\log(t_uw))^2\big)$ for zero-error data structures with word size $w$, worst-case update time $t_u$ and expected query time $t_q$. We extend this trade-off to amortized insertions and bounded error when a query may follow any update prefix. Our total-work theorem further gives a fixed distribution of insertions, each followed by a query, requiring $\Omega\big((\log n/\log\log n)^2\big)$ expected probes per operation (\cref{thm:total}).

% =====================================================================
\subsection{Previous work}\label{sec:previous}
% =====================================================================

\paragraph{Worst-case update time.}
Larsen, Weinstein and Yu \cite{larsen_crossing_2020} proved super-logarithmic lower bounds for Boolean dynamic problems, including range parity and Boolean polynomial evaluation, assuming worst-case update time. Their proof uses a one-way communication simulation and a peak-to-average lemma. The 2.5-round communication framework of \cite{ko_omega_2026,ko_omegalog_2026} reaches the quadratic-logarithmic scale by adding verification of the simulated query. We refer to those papers for the broader development of worst-case cell-probe lower bounds.

\paragraph{Conditional amortized lower bounds.}
Conditional lower bounds give polynomial restrictions on amortized update and query time, including for Boolean dynamic problems. Abboud and Vassilevska Williams \cite{abboud_popular_2014} obtained amortized update/query bounds under 3SUM, SETH, and APSP. Their examples include fully dynamic reachability and subgraph connectivity (3SUM), testing whether a digraph has more than two strongly connected components (SETH), and weighted $s$--$t$ shortest paths and maximum-weight bipartite matching (APSP), with hypothesis-dependent preprocessing restrictions. Henzinger, Krinninger, Nanongkai and Saranurak \cite{henzinger_unifying_2015} derived such bounds from the online Boolean matrix-vector multiplication (OMv) conjecture. For example, for fully dynamic subgraph connectivity on $n$ vertices, their conjecture rules out polynomial preprocessing together with amortized update time $O(n^{1-\eps})$ and amortized query time $O(n^{2-\eps})$, for any constant $\eps>0$.

However, recent developments have refuted the 3SUM, APSP, and SETH conjectures. An internal Claude research model discovered the algorithm yielding truly subquadratic 3SUM and truly subcubic APSP~\cite{alman_truly_2026}. Subexponential algorithms for 3-SAT by OpenAI\footnote{\emph{The Exponential Time Hypothesis is False}, OpenAI, October 25, 2026; \href{https://github.com/openai/math/blob/main/preprints/The-Exponential-Time-Hypothesis-is-False-October-25-2026/README.md}{paper and accompanying materials}.} and Claude (Anthropic)\footnote{\emph{A $2^{o(n)}$ Time Algorithm for 3-SAT}, Claude (Anthropic), 2026; \href{https://github.com/anthropics/formal-math/blob/main/eth/README.md}{paper and Lean formalization}.} refute ETH and hence SETH. The corresponding reductions remain correct, but none of these three hypotheses can now serve as evidence of hardness; see \cite{ko_omegalog_2026} for further discussion. These results do not refute the standard OMv conjecture used above. But they strengthen the motivation for unconditional amortized lower bounds, which we derive here.

\paragraph{Unconditional amortized cell-probe lower bounds.}
Amortized lower bounds charge the total work over a sequence of operations. Fredman and Saks \cite{fredman_cell_1989} already used alternating updates and queries, epochs placed backward from each query, and averaging over contiguous windows to control the work in these epochs. \Patrascu\ and Demaine \cite{patrascu_logarithmic_2006} proved randomized amortized $\Omega(\log n)$ lower bounds for partial sums and dynamic connectivity. Their information-transfer method uses interleaved updates and queries, forcing information from earlier updates to be recovered throughout the sequence.

Weinstein and Yu \cite{weinstein_amortized_2016} reached the quadratic-logarithmic scale for weighted two-dimensional orthogonal range counting. They gave interleaved sequences of $n/\polylog n$ updates and $n$ queries requiring total time $\Omega\big(n(\log n/\log\log n)^2\big)$. The bound holds even for data structures with success probability $\exp(-\widetilde\Omega(n))$. Their proof uses a four-party communication model. Bhattacharya, Henzinger and Neumann \cite{bhattacharya_new_2019} subsequently proved an $\Omega\big((\log n/\log\log n)^2\big)$ amortized lower bound for dynamic polynomial evaluation, and an amortized update/query trade-off for online matrix-vector multiplication, over polynomial-size finite fields. However, these results concern queries with $\Omega(\log n)$ output bits.

\paragraph{Super-logarithmic amortized bounds for Boolean problems.}
Weinstein and Yu \cite[Section~1.1]{weinstein_amortized_2016} observed that interleaving permits a zero-error amortized adaptation of Larsen's cell-sampling argument \cite{larsen_cell_2012}. The same approach also extends the Boolean polynomial-evaluation argument of Larsen, Weinstein and Yu \cite[Sections~4--5]{larsen_crossing_2020}, whose stated theorems assume worst-case updates. Over a field of size $\Theta(n^4)$, with logarithmic word size, this adaptation gives a fixed distribution of $n$ updates and $n$ interleaved queries requiring $\Omega\big(n\log^{3/2}n/(\log\log n)^2\big)$ expected probes from every zero-error randomized data structure.

To obtain this bound, update consecutive polynomial coefficients independently and uniformly, and place epochs backward from a random query time. Averaging bounds the work within and after the selected epoch. Alice sends an abort flag when these costs are too large. Bob conditions his posterior calculations on the entire first message, including this flag, before applying the peak-to-average lemma, so the cost cutoff causes only a constant multiplicative loss in advantage. Moving an epoch to a different interval of coefficients multiplies its polynomial contribution by $x^a$ for some $a\ge0$. At nonzero query points this preserves the independence of evaluations used in the communication lower bound. Thus the weaker super-logarithmic amortized bound follows from an extension of the LWY method. Zero error is used when consistency with the simulated memory fixes the correct answer. With bounded error, the inputs supporting that simulation can instead contain a disproportionate share of erroneous answers.

\paragraph{Amortization within the 2.5-round framework.}
We combine interleaving with the 2.5-round simulation and verification step of \cite{ko_omega_2026,ko_omegalog_2026} to obtain the quadratic-logarithmic amortized bound with constant per-query error. Epochs are placed backward from a random query time, and the classical averaging argument of Fredman and Saks \cite{fredman_cell_1989} controls their work (\cref{lem:window}). To handle errors, we additionally control the acceptance probability of each execution and ensure that every erroneous accepted disjoint answer requires an intersection within the selected epoch. Our error guarantee concerns each query separately; it differs from the exponentially small whole-sequence success probability allowed by Weinstein and Yu \cite{weinstein_amortized_2016}.

\subsection{Our results}
Our main result is a quadratic-logarithmic lower bound on the total work of randomized dynamic Disjointness with constant error. We also prove an update--query trade-off when a single query follows a sequence of updates whose length is not given in advance.

Throughout, fix the absolute constant $\delta=2^{-12}$ and assume that the data structure has error at most $\delta$ for each query of every fixed legal operation sequence. Preprocessing is unrestricted, and its probes are not counted. The model and error convention are specified in \cref{sec:model}.

\paragraph{Total work.}
We count all probes made by updates and queries, including writes made by queries. The following bound holds without any restriction on the cost of an individual operation.

\begin{theorem}[Total work]\label{thm:total}
Let $w=\Theta(\log N)$, $m=N$ and $n=N^{1/4}$. There is a distribution over inputs to multiphase Disjointness over $[N]$ with $m$ query sets, followed by $n$ distinct insertions each followed by a query, on which every randomized cell-probe data structure with the stated error guarantee makes $\Omega\big(n(\log N/\log\log N)^2\big)$ probes in expectation.
The expectations are over the distribution and the coins of the data structure, and the probes made by preprocessing are not counted.
\end{theorem}

The distribution in \cref{thm:total} is fixed independently of the data structure. Since the sequence has $2n$ operations, the theorem gives an expected average cost of $\Omega((\log N/\log\log N)^2)$ probes per operation.

\paragraph{Update--query trade-off.}
Now suppose that the data structure receives insertions followed by one final query, and is not told in advance when the insertions will stop. Its amortized insertion time is $t_u$ if, for every fixed input and every $u$, the first $u$ insertions make at most $ut_u$ probes in expectation over the coins. Its expected query time is $t_q$ if the query costs at most $t_q$ probes in expectation over the data structure's coins on every fixed input.

\begin{theorem}[Amortized Disjointness]\label{thm:disj}
Let $w=\Theta(\log N)$ and $m=N$. Every randomized cell-probe data structure with the stated error guarantee for insert-only multiphase Disjointness over the universe $[N]$, with $m$ query sets, amortized update time $1\le t_u=N^{o(1)}$ and expected query time $t_q$, has
\[ t_q=\Omega\!\left(\left(\frac{\log N}{\log(t_uw)}\right)^{2}\right). \]
In particular, $\max(t_u,t_q)=\Omega\big((\log N/\log\log N)^2\big)$. This holds already when every query set has $N^{1/4}$ elements and the inserted elements are distinct.
\end{theorem}

Compared with \cite{ko_omegalog_2026}, \cref{thm:disj} allows both amortized insertion time and constant error. The proof uses the assumption that the number of insertions is not given in advance; we leave open whether the same trade-off holds when the number of insertions is announced in advance.

\paragraph{Graph applications.}
The total-work bound transfers to incremental directed reachability, fully dynamic undirected triangle detection, and incremental subgraph connectivity (\cref{cor:graph-total}). Reachability is hard already on depth-two directed acyclic graphs with a fixed sink, and subgraph connectivity requires only vertex activations in a fixed undirected graph. The single-query trade-off also transfers to reachability and subgraph connectivity (\cref{cor:graph-query}). In \cref{sec:applications}, we prove these reductions and verify that they preserve the query schedule, the error guarantee, and the total probe count.

\subsection{Proof overview}\label{sec:overview}

Following \cite{ko_omega_2026,ko_omegalog_2026}, we use the Chronogram method to convert a fast data structure into a short 2.5-round communication protocol, then prove a communication lower bound. Interleaving and averaging over a random query time extend the simulation to amortized costs. To allow bounded error, we control the acceptance probability of each execution and the fraction of accepted answers that are wrong.

\paragraph{The Chronogram method.}
For the total-work bound, every update is followed by a query. This is the classical interleaving approach of \cite{fredman_cell_1989}, also emphasized by Weinstein and Yu \cite{weinstein_amortized_2016}: queries between updates prevent the data structure from postponing all its work until the end. The Chronogram accounting includes the probes and writes of these queries.

As in \cite{ko_omega_2026}, epochs grow geometrically backward in time. With amortized costs, a fixed epoch may receive much more work than its length times the average operation cost. We therefore place the epochs backward from a uniformly random query time $T$. Averaging over $T$ bounds each epoch's expected work by its share of the total work (\cref{lem:window}), using the amortized accounting of Fredman and Saks \cite[Section~4]{fredman_cell_1989}. Merlin aborts when an epoch exceeds a constant multiple of this bound, keeping the messages short. For the single-query trade-off, the same averaging applies to an update-only execution and a query simulated after a random prefix; each such query is legal because the number of updates is not given in advance.

The resulting 2.5-round protocol lets Bob verify Alice's simulation against the actual memory. Verification ensures that the simulated query produces the same answer as the actual query, though that answer may still be erroneous. Moreover, ordinary cell sampling can concentrate errors among the accepted executions. Bob therefore requires every probed cell from the selected epoch to lie in his sample and performs an additional independent random test. Its success probability depends on the number of distinct such cells, so that every execution satisfying the stated probe bounds is accepted with the same probability $\gamma=n^{-o(1)}$ before accounting for the negligible probability of exceeding the bound on sample size (\cref{lem:acceptance}). We keep the data structure's random tape random through the cost and error estimates. This lets us bound the fraction of accepted answers that are wrong.

\paragraph{The communication lower bound.}
In the zero-error argument, conditioning on accepted disjoint answers gives a distribution supported on disjoint pairs. With bounded error, this conditional distribution may contain intersecting pairs. We need a quantitative combinatorial bound that still rules it out when a small fraction of accepted answers is wrong.

The input distribution must be the same on every interval of a given length, since the epoch positions depend on $T$. The universe is $[n]\times[n]$, update $t$ inserts $(t,x_t)$ for a uniform $x_t\in[n]$, and every query set contains one independent uniform element $(t,S_q[t])$ for each $t\in[n]$. For an epoch of length $s$, our corruption bound states that two independent vectors in $[n]^s$, each with entropy at least $s\log n-s/100$, agree in some coordinate with probability at least $s/(4n)$ (\cref{lem:intersection}). We use $m\ge n^4$ query sets so that the correlation remaining after conditioning on acceptance is negligible compared with this probability. This accounts for the parameters $m=N=n^4$ and query-set size $n=N^{1/4}$ in \cref{thm:disj}.

Assume the per-query error guarantee with constant $\delta$. Alice declares the full input disjoint only when the simulated query returns disjoint and the sets are disjoint outside the selected epoch. An erroneous accepted $1$ then requires an intersection inside the epoch and none outside, an event of probability at most $s/n$. The error guarantee for each fixed input and the acceptance bound imply that such an error has probability at most $\gamma\delta s/n$.

A weighted averaging argument fixes the public coins while retaining enough accepted $1$ answers and conditional error at most a sufficiently small multiple of $s/n$. Conditioning on these answers yields high-entropy marginals with very little mutual information. Their product assigns probability less than $s/(4n)$ to intersecting pairs, contradicting the corruption bound (\cref{sec:disj-comm,sec:disj-error}). This communication lower bound allows a fraction $O(s/n)$ of accepted answers to be wrong, which suffices for constant error in the data structure.

\paragraph{Organization.}
\cref{sec:previous} discusses previous work, and \cref{sec:prelim} gives the model and communication game. After defining the hard distribution in \cref{sec:dist}, \cref{sec:outer} proves the simulation theorem and \cref{sec:inner} proves the communication lower bound. \Cref{sec:completing} combines them to obtain the dynamic bounds, and \cref{sec:applications} gives the graph applications.

% =====================================================================
\section{Preliminaries}\label{sec:prelim}
% =====================================================================

\paragraph{Notation.}
$[n]=\{1,\dots,n\}$. Logarithms are base $2$ unless written $\ln$. We omit floors and ceilings, rounding integer thresholds up when needed.
We write $\E_Y[f(Y)]$ for expectation over the distribution of $Y$;
multiple variables in the subscript indicate their joint distribution.
In $\E_Y[f(Y)|Z]$, this distribution is conditioned on $Z$.

\subsection{Information theory}\label{sec:it}

We follow the notation of \cite{ko_omega_2026}. $H(A)$ is the Shannon entropy of $A$, $H(A|B)$ the conditional entropy, and $I(A;B|C):=H(A|C)-H(A|B,C)$ the mutual information. For distributions $P$ and $Q$, $D(P\|Q):=\sum_xP(x)\log\frac{P(x)}{Q(x)}$ is the KL divergence, and $\norm{P-Q}_{TV}:=\frac12\norm{P-Q}_1=1-\sum_x\min\big(P(x),Q(x)\big)$ is the total variation distance. For an event $E$, $A|_E$ denotes the distribution of $A$ conditioned on $E$.

\begin{fact}[KL divergence and mutual information]\label[fact]{fact:kl-mi}
\[
 I(A;B|C)=\E_{(b,c)\sim(B,C)}\big[D\big(A|_{B=b,C=c}\,\big\|\,A|_{C=c}\big)\big].
\]
In particular, if $A$ is uniform on a set of size $M$, then
$I(A;B)=\E_{b\sim B}\big[\log M-H(A|B=b)\big]$.
\end{fact}

\begin{fact}[Pinsker's inequality]\label[fact]{fact:pinsker}
For distributions $P$ and $Q$, $\norm{P-Q}_{TV}\le\sqrt{D(P\|Q)/(2\log e)}$. In particular, for jointly distributed $(A,B)$,
\[ \norm{P_{AB}-P_A\otimes P_B}_{TV}\le\sqrt{I(A;B)/(2\log e)}. \]
\end{fact}

\begin{fact}[Chain rule]\label[fact]{fact:chain}
$I(A,D;B|C)=I(D;B|C)+I(A;B|C,D)$. Consequently, if $D$ is a function of $(A,C)$ or of $(B,C)$, then $I(A;B|C,D)\le I(A;B|C)$.
\end{fact}

\noindent We also use subadditivity, $H(V_1,\dots,V_s)\le\sum_jH(V_j)$, and data processing: $I(A';B'|C)\le I(A;B|C)$ if $A'$ is a function of $A$ and $B'$ is a function of $B$.

\begin{fact}[Entropy loss under conditioning]\label[fact]{fact:entropy-loss}
Let $A$ be uniform on a set of size $M$ and independent of $C$.
If $E$ is an event of probability $\rho>0$ and $W$ takes at most
$2^\lambda$ values, then
\[
 H(A|C,W,E)\ge\log M-\lambda-\log(1/\rho),
\]
where the entropy is evaluated under the distribution conditioned on $E$.
\end{fact}
\begin{proof}
Let $P$ be the distribution of $(A,C)$ and $P_E$ its distribution conditioned on $E$.
The likelihood ratio is
$P_E(a,c)/P(a,c)=\Pr[E|A=a,C=c]/\rho\le1/\rho$.
Since $A$ is uniform and independent of $C$, the KL chain rule gives
\[
 \log M-H(A|C,E)\le D(P_E\|P)\le\log(1/\rho).
\]
Revealing $W$ costs at most $H(W|C,E)\le\lambda$ further bits.
\end{proof}

\noindent Finally, we use Markov's inequality and the Chernoff bound $\Pr[Y\ge3\mu/2]\le e^{-\mu/12}$, which holds for a sum $Y$ of independent indicators and every $\mu\ge\E_Y[Y]$.

\subsection{Cell-probe model}\label{sec:model}

Memory consists of cells of $w$ bits, and an address fits in one cell.
A data structure is charged one probe for each cell it reads or writes,
and computation is free. All information kept between operations,
including any random state, is stored in the cells. Preprocessing is
unrestricted, and its probes are not counted.

A randomized data structure uses coins independent of the input, and
preprocessing and every operation terminate with probability $1$ on
every legal input. We give preprocessing and each operation position its
own independent coins; this does not change the distribution of any
execution.

\paragraph{Error convention.}
For every fixed preprocessing input and every fixed legal operation
sequence, each query has error probability at most $\delta$ over the data
structure's coins. This guarantee applies to each query separately;
we do not require all answers to be correct simultaneously. Our proofs
use this guarantee, and the time bounds below, only for sequences whose
inserted elements are distinct.

\paragraph{Multiphase Disjointness.}
In insert-only multiphase Disjointness \cite{patrascu_mihai_towards_2010,ko_omegalog_2026}, the data structure maintains a set $X\subseteq[N]$, initially empty.
\begin{enumerate}[label=(\roman*)]
\item \emph{Phase I.} Sets $S_1,\dots,S_m\subseteq[N]$ arrive and are preprocessed.
\item \emph{Phase II.} Elements are inserted into $X$ one at a time.
\item \emph{Phase III.} An index $q\in[m]$ arrives, and the data structure outputs $\DISJ(S_q,X):=\mathbf 1[S_q\cap X=\emptyset]$.
\end{enumerate}

\paragraph{Time.}
In \cref{thm:disj}, a single query may follow any sequence of updates;
the number of updates is not given to the data structure in advance.
Amortized update time $t_u$ means that, for every fixed preprocessing
input and every fixed sequence of $u$ updates, those updates make at
most $ut_u$ probes in expectation. Expected query time $t_q$ means that
every query following such a sequence makes at most $t_q$ probes in
expectation. Both expectations are over the data structure's coins.
The update bound has no additive term depending on $N$.

In \cref{thm:total}, every update is followed by a query; queries may write cells, and we count all probes made after preprocessing.

\subsection{The 2.5-round game}\label{sec:game}

We use the 2.5-round Multiphase Communication Game of \cite[Protocol~3]{ko_omega_2026}. Alice holds $\vS=(S_1,\dots,S_m)$, Bob holds $X$, and Merlin holds both. Merlin sends Bob a message $U$, which is a function of $(\vS,X)$. Bob sends Alice a message $\Phi$, which depends on $U$, $X$ and Bob's private coins, but not on the query. Then a query index $Q\in[m]$ is given to Alice only, and she sends Bob a message $A_Q$, where $A_q$ is a function of $(\vS,\Phi,q)$ for every $q$. Finally, Bob either accepts and outputs an answer, or outputs $\FAIL$, as a function of $U$, $A_Q$ and his private coins. Messages are padded to fixed lengths. As in \cite{ko_omegalog_2026}, Bob outputs $\FAIL$ when his verification fails, where \cite{ko_omega_2026} lets him output a random guess. In our protocols, the players also share public coins, independent of the input, which include the data structure's random tape and are fixed by averaging only after the cost and error estimates. Merlin's message is prescribed by the protocol; he is not an adversarial prover in this game.

% =====================================================================
\section{Main proof}\label{sec:proof}
% =====================================================================

Following \cite{ko_omega_2026,ko_omegalog_2026}, the proof has four steps.
Step~1 shows that a fast data structure yields a too-good 2.5-round
protocol, whose accepted answers $1$ are rarely wrong. Conditioning on
these answers gives a near-rectangle that is nearly monochromatic
(Step~2). The product of its marginals is a nearly monochromatic
rectangle (Step~3), and a corruption bound rules this out (Step~4).

\subsection{The hard distribution and operation sequences}\label{sec:dist}
Let $n$ be the number of updates in the operation sequence.
The universe is $[n]\times[n]$.
Independently sample each $S_q\in[n]^n$ and $X\in[n]^n$ uniformly.
The query set represented by $S_q$ is
$\{(t,S_q[t]):t\in[n]\}$, and update $t$ inserts $(t,x_t)$.
We identify $S_q$ and $X$ with the sets they represent. These sets
intersect exactly when $S_q[t]=x_t$ for some $t$.
For independent uniform $Q\in[m]$,
\begin{equation}\label{eq:yes}
 \Pr[\DISJ(S_Q,X)=1]=(1-1/n)^n\ge1/4\qquad(n\ge2).
\end{equation}

\paragraph{Single-query setting.}
Execute all $n$ updates without queries, and let $c_t$ be the cost of
update $t$. Separately simulate a query on an independent uniform index
$Q_t$ after each update $t$, using the coins of operation position
$t+1$, as an actual query after that prefix would, and let $\lambda_t$
be its cost. Discard its writes
before continuing the updates. Since the number of updates is not given
in advance, this query has the same distribution as on that legal
shorter input.

\paragraph{Interleaved setting.}
After update $t$, execute a query on an independent uniform index $Q_t$.
Let $c_t$ be the combined cost of update and query $t$, and let
$\lambda_t\le c_t$ be the query cost. All writes are retained.
In both settings, write $\mathbf Q=(Q_1,\ldots,Q_n)$ for the
independent uniform query indices (simulated separately in the single-query setting).
The data structure's tape $R$ is independent of the inputs and $\mathbf Q$.

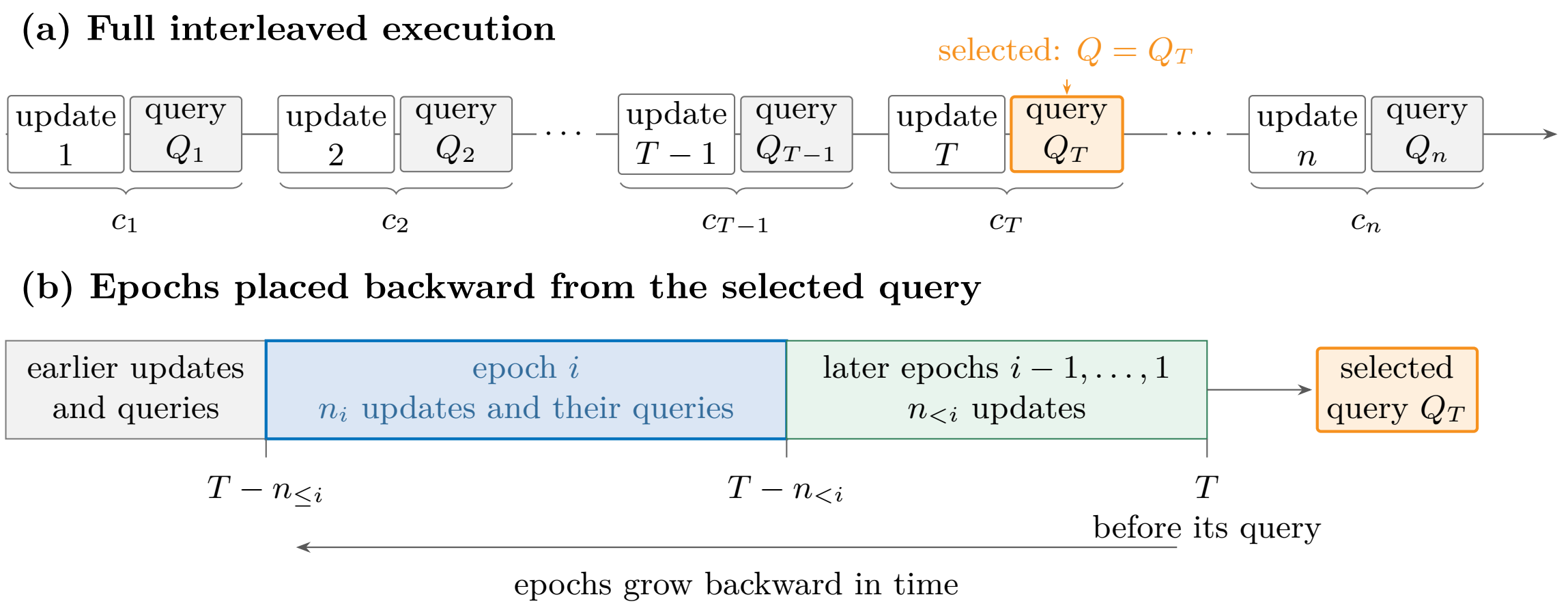
\begin{figure}[htbp]
\centering
\begin{tikzpicture}[
  x=0.97cm,y=1cm,font=\footnotesize,
  operation/.style={draw=black!55,fill=white,rounded corners=1pt,
    minimum width=1.08cm,minimum height=0.72cm,align=center,inner sep=2pt},
  query/.style={operation,fill=black!5},
  selected/.style={query,draw=BurntOrange,fill=BurntOrange!12,thick},
  pairbrace/.style={decorate,decoration={brace,mirror,amplitude=3pt},
    draw=black!55},
  timearrow/.style={-{Stealth[length=4pt]},draw=black!65}
]
% The full execution is fixed before the independent choice of T.
\node[anchor=west,font=\small\bfseries] at (0,0.35)
  {(a) Full interleaved execution};
\draw[timearrow] (0,-0.65) -- (15.5,-0.65);
\node[operation] at (0.6,-0.65) {update\\$1$};
\node[query] at (1.8,-0.65) {query\\$Q_1$};
\node[operation] at (3.3,-0.65) {update\\$2$};
\node[query] at (4.5,-0.65) {query\\$Q_2$};
\node[fill=white,inner sep=2pt] at (5.6,-0.65) {$\cdots$};
\node[operation] at (6.7,-0.65) {update\\$T-1$};
\node[query] at (7.9,-0.65) {query\\$Q_{T-1}$};
\node[operation] at (9.4,-0.65) {update\\$T$};
\node[selected] at (10.6,-0.65) {query\\$Q_T$};
\node[fill=white,inner sep=2pt] at (11.9,-0.65) {$\cdots$};
\node[operation] at (13.0,-0.65) {update\\$n$};
\node[query] at (14.2,-0.65) {query\\$Q_n$};
\node[text=BurntOrange,anchor=south] at (10.6,-0.13)
  {selected: $Q=Q_T$};
\draw[BurntOrange,-{Stealth[length=3pt]}]
  (10.6,-0.12) -- (10.6,-0.27);
\draw[pairbrace] (0.04,-1.12) -- (2.36,-1.12)
  node[midway,below=5pt] {$c_1$};
\draw[pairbrace] (2.74,-1.12) -- (5.06,-1.12)
  node[midway,below=5pt] {$c_2$};
\draw[pairbrace] (6.14,-1.12) -- (8.46,-1.12)
  node[midway,below=5pt] {$c_{T-1}$};
\draw[pairbrace] (8.84,-1.12) -- (11.16,-1.12)
  node[midway,below=5pt] {$c_T$};
\draw[pairbrace] (12.44,-1.12) -- (14.76,-1.12)
  node[midway,below=5pt] {$c_n$};

% Boundaries are after the indicated update--query pair, except at T.
\node[anchor=west,font=\small\bfseries] at (0,-2.15)
  {(b) Epochs placed backward from the selected query};
\filldraw[fill=black!5,draw=black!50]
  (0,-2.65) rectangle (2.6,-3.6);
\filldraw[fill=RoyalBlue!12,draw=RoyalBlue,thick]
  (2.6,-2.65) rectangle (7.8,-3.6);
\filldraw[fill=ForestGreen!8,draw=ForestGreen!70!black]
  (7.8,-2.65) rectangle (12,-3.6);
\node[align=center] at (1.3,-3.125) {earlier updates\\and queries};
\node[align=center,text=RoyalBlue!70!black] at (5.2,-3.125)
  {epoch $i$\\$n_i$ updates and their queries};
\node[align=center] at (9.9,-3.125)
  {later epochs $i-1,\ldots,1$\\$n_{<i}$ updates};
\draw[timearrow] (12,-3.125) -- (13.05,-3.125);
\node[selected,minimum width=1.55cm] at (13.9,-3.125)
  {selected\\query $Q_T$};
\foreach \x in {2.6,7.8,12}
  \draw[black!60] (\x,-3.6) -- (\x,-3.78);
\node[anchor=north] at (2.6,-3.82) {$T-n_{\le i}$};
\node[anchor=north] at (7.8,-3.82) {$T-n_{<i}$};
\node[anchor=north,align=center] at (12,-3.82)
  {$T$\\before its query};
\draw[timearrow] (11.7,-4.65) -- (2.9,-4.65)
  node[midway,below=3pt] {epochs grow backward in time};
\end{tikzpicture}
\caption{Interleaving and the random query time (schematic; not to scale).
The full execution alternates updates and queries, with pair cost $c_t$.
Independently choose $T$ uniformly from $\{n/2,\ldots,n\}$ and select
the query with index $Q=Q_T$.
Epoch $i$ contains updates $T-n_{\le i}+1,\ldots,T-n_{<i}$ and their
following queries. Earlier query writes are retained.
Epoch~$1$ includes update $T$ but excludes the selected query;
$c_T$ counts both. Operations after $T$ remain part of the full execution
and its total cost, but are not simulated.}
\label{fig:interleaving}
\end{figure}

Choose $T$ uniformly from $\{n/2,\ldots,n\}$, independently
of the inputs, query indices, and random tape. The selected query is the query after update $T$,
with index $Q$. The cost vector $(c_1,\ldots,c_n)$ is defined before and
independently of $T$. When the interleaved protocol is formed, $T$ and all
query indices except $Q_T$ are public; the remaining index is uniform
and independent of them. Operations after $T$ are never simulated.
\Cref{fig:interleaving} illustrates this timeline and the backward epoch
placement defined below.

\subsection{Step 1: a too-good 2.5-round protocol}\label{sec:outer}

\subsubsection{Epochs before a random query time}\label{sec:epochs}
Suppose, for the chosen setting, that parameters $a,b\ge1$ satisfy
\begin{equation}\label{eq:costs}
 \E_{\vS,X,\mathbf Q,R}\!\left[\sum_{t=1}^n c_t\right]\le na,
 \qquad \E_{\vS,X,\mathbf Q,R,T}[\lambda_T]\le b.
\end{equation}
Here the expectations are over the query sets $\vS$, update vector $X$,
query indices $\mathbf Q$, and data structure's tape $R$; the second
also averages over the independent uniform query time $T$.
In the single-query setting the costs $c_t$ do not depend on $\mathbf Q$.
The tape $R$ remains random through the error estimates.
Let $\beta$ be $(aw)^2$ rounded
up to a power of two, and set
\[
 n_i=\beta^i,\quad
 n_{<i}=\sum_{j<i}n_j,\quad n_{\le i}=n_i+n_{<i}.
\]
Let $\ell$ be maximal with $n_{\le\ell}\le n/2$,
and let $\cI=\{i\in[\ell]:i\ge2(\ell+1)/3\}$ and $k=|\cI|$.
Epoch $i$ consists of updates $T-n_{\le i}+1,\ldots,T-n_{<i}$ and,
in the interleaved setting, their following queries except the
selected query at $T$. Write $Y^{(i)}$ for coordinates in that epoch,
$Y^{(>i)}$ for coordinates before it, and $Y^{(\ne i)}$ for all
coordinates outside it, including those after $T$.

For $a=n^{o(1)}$ and $w=\Theta(\log n)$, the geometric series gives
\begin{equation}\label{eq:params}
 k=\Theta\!\left(\frac{\log n}{\log(aw)}\right),\qquad
 n_i\ge n^{2/3}/3\ (i\in\cI),\qquad n_{<i}\le2n_i/\beta.
\end{equation}
Indeed, maximality gives
$\beta^{\ell+1}>n/4$; raising to power $2/3$ gives the middle inequality.

The following elementary averaging bound supplies the amortized Chronogram
accounting \cite[Lemma~1]{fredman_cell_1989}.

\begin{lemma}[Averaging over query times]\label[lemma]{lem:window}
Let $T$ be uniform on $\{n/2,\ldots,n\}$.
For fixed nonnegative $c_1,\ldots,c_n$ and integers $d\ge0,h\ge1$ with
$d+h\le n/2$,
\[
 \E_T\!\left[\sum_{t=T-d-h+1}^{T-d}c_t\right]
 \le\frac{2h}{n}\sum_t c_t.
\]
For a random cost vector $\mathbf c=(c_1,\ldots,c_n)$ independent of $T$,
averaging also over $\mathbf c$ gives
\[
 \E_{\mathbf c,T}\!\left[\sum_{t=T-d-h+1}^{T-d}c_t\right]
 \le\frac{2h}{n}\E_{\mathbf c}\!\left[\sum_t c_t\right].
\]
\end{lemma}
\begin{proof}
Each position belongs to at most $h$ windows, and each possible $T$
has probability at most $2/n$. Every window lies in $[n]$.
\end{proof}

Throughout the simulation, assume for contradiction that
\begin{equation}\label{eq:fast}
 a=n^{o(1)},\quad w=\Theta(\log n),\quad
 b=o\!\left(\left(\frac{\log n}{\log(aw)}\right)^2\right).
\end{equation}

\subsubsection{Simulation theorem}\label{sec:step1}

The following theorem is Step~1 of the argument. It gives the
too-good protocol that \cref{sec:inner} rules out.

\begin{theorem}[Simulation theorem]\label{thm:simulation}
Assume \eqref{eq:costs} and \eqref{eq:fast}, and suppose the data
structure satisfies the error guarantee of \cref{sec:model} with $\delta=2^{-12}$.
There are fixed public coins, including an epoch $i\in\cI$ of length
$s=n_i$, for which the protocol below has
\[
 |U|\le n^{5/4},\qquad \Phi=(X^{(\ne i)},\Psi),\quad
 |\Psi|\le s/200,\qquad |A_q|\le\log^3n.
\]
The event $B$ of accepting with output $1$ is determined by
$(U,\Phi,A_Q)$ and satisfies
\[
 \rho:=\Pr[B]\ge n^{-o(1)},\qquad
 \Pr[\exists j\in\text{epoch }i:S_Q[j]=x_j|B]\le\frac{s}{64n}.
\]
The input distribution remains the product distribution of
\cref{sec:dist}.
\end{theorem}

The public coins are $R,T,i$, with $i$ uniform in $\cI$, and the
other query indices in the interleaved setting. Bob uses private
coins for sampling and the additional random test. Fix the public coins only to describe
the protocol; they remain random in the cost and error estimates.

\paragraph{Memory.}
Alice reconstructs the memory $M_{>i}$ before epoch $i$ from
$\vS,X^{(>i)}$, the random tape $R$, and earlier query indices.
Let $U_j$ record the distinct cells written in epoch $j$ and their
contents at its end. Let $C_j$ consist of the cells of $U_j$ not written
again after epoch $j$ and before the selected query, and put
$C_{<i}=\bigcup_{j<i}C_j$.
The memory at the selected query differs from $M_{>i}$ only on
$C_i\cup C_{<i}$.
Write $c^{(j)}$ for the number of probes in epoch $j$, excluding the
selected query, and set $c^{(<i)}=\sum_{j<i}c^{(j)}$. Then
\[
 |C_i|\le c^{(i)},\qquad |C_{<i}|\le c^{(<i)}.
\]

\paragraph{Selected query.}
Simulate the query while storing locally the most recent value of
each cell read or written. Reads use the local value when available
and otherwise probe the memory before the query; writes only change
the local value. This gives the same answer using no more probes.
Let $L$ count these memory probes and let $d_j$ count those to $C_j$.
Each cell is read from memory at most once, so
\[
 \sum_{j\in\cI}d_j\le L\le\lambda_T.
\]
Simulations that exceed a sufficiently large computation limit are
rejected.\footnote{For almost every fixed random tape $R$, all legal query
computations considered here terminate. There are finitely many at
each input size, so choose a limit exceeding all their running times.
This rejects only inconsistent simulations. The public coins fixed in
\cref{sec:disj-error} can be chosen outside the null set of other tapes.}

\paragraph{Parameters.}
Let $p$ be the sampling probability, and let $L_*$ and $r$ bound the
total query probes and those to $C_i$, respectively. Set
\begin{equation}\label{eq:sample-params}
 \begin{gathered}
 K=32,\qquad p=\frac1{2\beta}=2^{-h},\qquad L_*=Kb,\\
 r=\frac{Kb}{k},\qquad \gamma=p^r,\qquad \sigma=2Kpa n_i.
 \end{gathered}
\end{equation}
By \cref{eq:params,eq:fast}, $hr=o(\log n)$,
$\gamma=n^{-o(1)}$, and $\sigma\ge n^{2/3-o(1)}$ uniformly over
$i\in\cI$; hence $r\le\sigma/2$.

\paragraph{Protocol.}
All messages are padded. An abort before Alice's simulation is flagged
by $\Psi=\bot$: Bob still sends $\Phi=(X^{(\ne i)},\Psi)$, Alice
sends a failure symbol, and Bob outputs $\FAIL$.
\begin{enumerate}[leftmargin=*,label=\arabic*.]
\item \emph{Merlin to Bob.} Send $U=\bot$ and abort if
$c^{(i)}>2Ka n_i$ or $c^{(<i)}>2Ka n_{<i}$.
Otherwise send $U=(U_i,\ldots,U_1)$ with delimiters. Bob recovers
$C_i,C_{<i}$ and all their contents before the selected query.
\item \emph{Bob to Alice.} Independently sample each cell of $C_i$
with probability $p$ to obtain $\widetilde C_i$. Abort if
$|\widetilde C_i|>2\sigma$. Independently draw a string
$F\in\{0,1\}^{hr}$. Send
$\Phi=(X^{(\ne i)},\Psi)$, where $\Psi$ contains the addresses and
contents of $C_{<i}\cup\widetilde C_i$, its delimiters, and $F$.
An abort is explicitly flagged. Alice ignores $F$ during simulation.
\item \emph{Alice to Bob.} Given $Q$, reconstruct $M_{>i}$ and simulate
the query as described above. For each read not answered from local
storage, use the listed cell contents when available, and otherwise
use $M_{>i}$. Stop if more than $L_*$ such reads would be required.
Send their address--content pairs and the output $z_Q$ defined below,
or a failure symbol.
\item \emph{Verification.} Reject an unfinished run.
Check the reported contents of every probed cell in $C_i\cup C_{<i}$
against $U$. Require that every probed address in $C_i$ was sampled.
Let $d$ be the number of distinct such addresses, and reject if $d>r$.
Finally accept exactly when the first $h(r-d)$ bits of $F$ are zero.
On acceptance output $z_Q$; otherwise output $\FAIL$.
\end{enumerate}

If the simulated answer to the query after update $T$ is $\widehat y_T$, Alice obtains an
answer for the full input by also checking that the sets are disjoint outside epoch $i$:
\begin{equation}\label{eq:disj-filter}
 z_Q=\widehat y_T\wedge
 \mathbf1[S_Q[j]\ne x_j\text{ for every }j\notin\text{epoch }i].
\end{equation}
This uses only Alice's input and the already transmitted $X^{(\ne i)}$.
When $\widehat y_T$ is correct, $z_Q$ is the full-input answer.

\begin{lemma}[Message lengths and verification]\label[lemma]{lem:messages}
Under \eqref{eq:fast}, the protocol has
\[
 |U|=O(awn),\qquad |\Psi|\le n_i/200,\qquad
 |A_q|=O(wb+\log n)=o(\log^3n).
\]
Its acceptance decision and output are functions of $(U,\Phi,A_Q)$.
In every accepted simulation, each read returns the value it would
return in the actual selected query. In particular, the simulated answer
$\widehat y_T$ equals the data structure's answer, though that answer
may be erroneous.
\end{lemma}
\begin{proof}
The cells in $C_{<i}$ number at most
$2Ka n_{<i}\le4Ka n_i/\beta$, and the sampled cells number at most
$2\sigma=2Ka n_i/\beta$. Each address--content pair has $2w$ bits.
Thus the cell part of $\Psi$ has at most
$12Ka w n_i/\beta=O(n_i/(aw))=o(n_i)$ bits.
Delimiters and flags cost $O(\log n)$, as does the string $F$.
The total is at most $n_i/200$.
When Merlin does not abort, the lists in $U$ contain at most
$c^{(i)}+c^{(<i)}\le2Ka(n_i+n_{<i})$ pairs;
epoch delimiters add $O(\log^2n)$ bits. This gives $|U|=O(awn)$.
Alice records at most $L_*$ address--content pairs plus flags and a bit.

The sample and the string $F$ are recorded in $\Psi$, so Bob's
decision is determined by $(U,\Phi,A_Q)$. Equivalently, he uses only
$U,A_Q$ and his private coins, meeting the restriction on Bob's output
of \cite[Protocol~3]{ko_omega_2026}. Every accepted read has either
its verified current content or its unchanged content in $M_{>i}$.
Induction on the simulated operations, including local reads and
writes, shows that the simulated answer equals the actual query answer.
\end{proof}

\begin{lemma}[Acceptance probability]\label[lemma]{lem:acceptance}
Let $\cE$ be the event that Merlin's two probe bounds hold and the query
simulation using the actual memory has $L\le L_*$ and $d_i\le r$.
Over inputs and all public randomness, $\Pr[\cE]\ge7/8$.
For every fixed input, choice of public coins, and selected query, the
acceptance probability over Bob's private coins is zero outside $\cE$ and $\gamma(1-\eta)$ on $\cE$, where
$0\le\eta\le2^{-\sqrt n}$ depends on the execution.
In particular acceptance is at most $\gamma$ on every execution.
\end{lemma}
\begin{proof}
By \cref{lem:window} and \eqref{eq:costs},
for every fixed $i\in\cI$,
\[
 \E_{\vS,X,\mathbf Q,R,T}[c^{(i)}]\le2an_i,
 \qquad \E_{\vS,X,\mathbf Q,R,T}[c^{(<i)}]\le2an_{<i}.
\]
In the interleaved setting the excluded selected query only reduces
these epoch costs. Further,
\[
 \E_{\vS,X,\mathbf Q,R,T}[L]\le b,
 \qquad \E_{\vS,X,\mathbf Q,R,T,i}[d_i]\le b/k,
\]
where the latter expectation also averages over an independent uniform
$i\in\cI$.
Markov's inequality and the union bound give total failure probability
at most $4/K=1/8$.

Fix an execution in $\cE$ with $d=d_i$.
All $d$ cells of $C_i$ probed by the simulation are sampled with probability $p^d$.
On this event, Alice's simulation uses the same values as the actual
query. Conversely, \cref{lem:messages} and Bob's check that every probed
cell in $C_i$ was sampled make this event necessary for acceptance.
The independent test on $F$ succeeds with probability $p^{r-d}$.
Ignoring the bound on sample size, the acceptance probability is
therefore $p^d p^{r-d}=\gamma$.

Conditioned on these $d$ cells being sampled, the sample size is $d$
plus a sum $Y$ of independent indicators with $\E[Y]\le pc^{(i)}\le\sigma$.
Because $d\le r\le\sigma/2$, the Chernoff bound gives
\[
 \Pr\big[\abs{\Csam}>2\sigma\,\big|\,\text{all $d$ cells are sampled}\big]
 \le\Pr[Y\ge3\sigma/2]\le e^{-\sigma/12}\le2^{-\sqrt n}.
\]
The string $F$ is independent of sampling. Outside $\cE$, one of the stated probe bounds is violated;
\cref{lem:messages} therefore rules out acceptance.
\end{proof}

\subsubsection{Completing the simulation theorem}\label{sec:disj-error}

\begin{proof}[Proof of \cref{thm:simulation}]
For a fixed epoch of length $s=n_i$, the test \eqref{eq:disj-filter}
forces any erroneous reported $1$ to occur on the input event
\[
 D_i=\{\text{the sets are disjoint outside epoch }i,
       \text{ but intersect within epoch }i\}.
\]
For every fixed $T,i$, the independent uniform coordinates give
\begin{equation}\label{eq:rare}
 \Pr[D_i]=(1-1/n)^{n-s}\big(1-(1-1/n)^s\big)\le s/n.
\end{equation}
On an input in $D_i$ the correct answer to the query after update $T$ is $0$, since
the entire epoch precedes the selected query. Thus a reported $1$
also requires error in that query.

For each fixed full input and fixed query sequence, that error has
probability at most $\delta$ over $R$. Apply this error guarantee
before fixing $R$. Since \cref{lem:acceptance} bounds acceptance by
$\gamma$ for every realization of $R$ and the input, for fixed $T,i$
and fixed other query indices we obtain
\begin{equation}\label{eq:localized-error}
 \Pr[\text{accepted erroneous }1]\le\gamma\delta\Pr[D_i]
 \le\gamma\delta s/n.
\end{equation}
The factor $s/n$ uses the error guarantee for each fixed input in $D_i$,
before the random tape is fixed.

Let $\omega$ be all public randomness, including $R,T,i$ and the
other query indices. Let $\rho_\omega$ be the probability, over the
unchanged input distribution and Bob's private coins, of an accepted
output $1$ in the resulting fixed protocol. Let $e_\omega$ be the
probability of an accepted erroneous $1$, and let $s_\omega=n_i$.
By \eqref{eq:yes} and \cref{lem:acceptance},
\begin{equation}\label{eq:rho-average}
 \E_\omega[\rho_\omega]
 \ge\gamma(1-2^{-\sqrt n})(1/4-1/8-\delta)\ge\gamma/16.
\end{equation}
Here we count disjoint inputs for which $\cE$ holds and the query is
correct; on those inputs the sets are also disjoint outside epoch $i$.
By \eqref{eq:localized-error},
\[
 \E_\omega[(n/s_\omega)e_\omega]\le\gamma\delta.
\]
Consequently
\[
 \E_\omega[\rho_\omega-64(n/s_\omega)e_\omega]
 \ge\gamma(1/16-64\delta)=3\gamma/64>\gamma/32.
\]
Fix public coins for which the expression inside the expectation is at
least $\gamma/32$. Then
\begin{equation}\label{eq:fixed-quality}
 \rho\ge\gamma/32=n^{-o(1)},\qquad e/\rho\le s/(64n).
\end{equation}
This fixes the public coins without conditioning the input distribution;
the fixed tape need not be correct on every input.
Every accepted $1$ on sets intersecting within epoch $i$ is erroneous, so
\eqref{eq:fixed-quality} gives the claimed conditional intersection bound.
The message bounds follow from \cref{lem:messages} and \eqref{eq:fast}.
\end{proof}

\subsection{Communication lower bound (Steps 2--4)}
\label{sec:inner}\label{sec:disj-comm}

For $m\ge n^4$, we rule out the protocol supplied by
\cref{thm:simulation}. The argument depends only on the message bounds and input
distribution.

\begin{lemma}[Communication lower bound]\label[lemma]{lem:robust}
Fix an epoch of length
$n^{2/3}/3\le s\le n$. Let $m\ge n^4$ and consider a 2.5-round
protocol (\cref{sec:game}) with fixed public coins on the distribution of \cref{sec:dist},
satisfying
\[
 |U|\le n^{5/4},\qquad \Phi=(X^{(\ne i)},\Psi),\quad
 |\Psi|\le s/200,\qquad |A_q|\le\log^3n.
\]
Suppose the event $B$ of accepting with output $1$ is determined by
$(U,\Phi,A_Q)$, and $\rho=\Pr[B]\ge n^{-1/2}$. Then
\[
 \Pr[\exists j\in\text{epoch }i:S_Q[j]=x_j|B]>\frac{s}{64n}.
\]
\end{lemma}

Suppose, for contradiction, that the conditional intersection probability
is at most $s/(64n)$. As in \cite{ko_omegalog_2026}, we call a
distribution of $(S_Q^{(i)},X^{(i)})$ a \emph{near-rectangle} if both
marginals have high entropy and the two sides have small mutual
information, a \emph{rectangle} if it is a product distribution, and
\emph{nearly monochromatic} if it puts small mass on intersecting pairs.

\subsubsection{Step 2: a nearly monochromatic near-rectangle}
Let $\Aall:=(A_1,\ldots,A_m)$, $S_{<q}=(S_1,\ldots,S_{q-1})$, and set 
\[
 Z':=(\Phi,\Aall,S_{<Q},Q),\qquad Z:=(Z',\mathbf1_B).
\]
For $z$ in the support of $Z$, let $\mu_z$ be the distribution of
$(S_Q^{(i)},X^{(i)})$ conditioned on $Z=z$, with marginals
$\mu_z^S,\mu_z^X$. Here $\mathbf1_B$ indicates acceptance with output $1$;
conditioning on the event $B$ means restricting to $\mathbf1_B=1$.

\begin{claim}\label[claim]{clm:conditioning}
There is a $z$ in the support of $Z|B$ for which
\[
 H(\mu_z^S),H(\mu_z^X)\ge s\log n-s/100,\qquad
 I(S_Q^{(i)};X^{(i)}|Z=z)\le16n^{-9/4},
\]
and $\Pr_{\mu_z}[\exists j\in\text{epoch }i:S_Q[j]=x_j]\le s/(8n)$.
\end{claim}
\begin{proof}
The index $Q$ is uniform and independent of $(\vS,X,U,\Phi,\Aall)$,
$\Aall$ is a function of $(\vS,\Phi)$, and $\mathbf1_B$ is a function of
$(Z',U)$. The chain rule gives
\begin{align}
 I(S_Q;X^{(i)},U|Z')
 &=\frac1m I(\vS;X^{(i)},U|\Phi,\Aall)\notag\\
 &\le\frac1m I(\vS;X^{(i)},U|\Phi)
 \le\frac{H(U)+H(\Psi)}m\le\frac{2n^{5/4}}m.\label{eq:before-mi}
\end{align}
For the penultimate inequality, $\Phi=(X^{(\ne i)},\Psi)$ and independence
of $\vS$ and $X$ imply
\[
 I(\vS;X^{(i)},U|X^{(\ne i)},\Psi)
 \le I(\vS;X^{(i)},U,\Psi|X^{(\ne i)})
 \le H(U)+H(\Psi).
\]
Likewise, independence of $S_1,\ldots,S_m$ and $X$ gives
\begin{align}
 I(S_Q;Z',U)
 &=\frac1m I(\vS;\Phi,\Aall,U)\notag\\
 &\le\frac{H(\Psi)+H(\Aall)+H(U)}m
 \le\log^3n+1.\label{eq:query-info}
\end{align}

For the entropy bounds, define the nonnegative Shannon entropy deficits
\[
 \Delta_S(z):=s\log n-H(\mu_z^S),\qquad
 \Delta_X(z):=s\log n-H(\mu_z^X).
\]
Since $S_Q^{(i)}$ is uniform on $[n]^s$ and $Z$ is a function of
$(Z',U)$, \eqref{eq:query-info} gives
$\E_{z\sim Z}[\Delta_S(z)]=I(S_Q^{(i)};Z)\le\log^3n+1$.
Restricting this nonnegative average to $B$ yields
\begin{equation}\label{eq:query-entropy}
 \E_{z\sim Z|B}[\Delta_S(z)]\le(\log^3n+1)/\rho.
\end{equation}

For the marginal distribution of $X^{(i)}$, put $Y=(\vS,X^{(\ne i)},Q)$.
The vector $X^{(i)}$ is uniform on $[n]^s$ and independent of $Y$.
Apply \cref{fact:entropy-loss} with the event $B$ and message $\Psi$:
\[
 H(X^{(i)}|Y,\Psi,B)\ge s\log n-s/200-\log(1/\rho).
\]
Since $Z'$ is a function of $(Y,\Psi)$, discarding this extra
conditioning can only increase entropy. Thus
\begin{equation}\label{eq:update-entropy}
 \E_{z\sim Z|B}[\Delta_X(z)]
 =s\log n-H(X^{(i)}|Z',B)
 \le s/200+\log(1/\rho).
\end{equation}
Combining \cref{eq:query-entropy,eq:update-entropy}, and using
$s\ge n^{2/3}/3$ and $\rho\ge n^{-1/2}$, gives
\[
 \E_{z\sim Z|B}[\Delta_S(z)+\Delta_X(z)]
 \le s/200+(\log^3n+1)/\rho+\log(1/\rho)=s/200+o(s).
\]
Markov's inequality therefore bounds the probability that their sum
exceeds $s/100$ by $1/2+o(1)$. Otherwise both marginal entropy
bounds in the claim hold.

For the mutual information, put $W=(X^{(i)},U)$.
Because $\mathbf1_B$ is a function of $(Z',W)$,
the chain rule, data processing, and \eqref{eq:before-mi} imply
\[
 I(S_Q^{(i)};X^{(i)}|Z)
 \le I(S_Q;W|Z',\mathbf1_B)
 \le I(S_Q;W|Z')\le2n^{5/4}/m.
\]
This averages over both values of the acceptance indicator.
Restricting to $B$ gives
\[
 \E_{z\sim Z|B}\!\left[I(S_Q^{(i)};X^{(i)}|Z=z)\right]
 \le2n^{5/4}/(m\rho).
\]
Markov's inequality then shows that
\begin{equation}\label{eq:near-product}
 I(S_Q^{(i)};X^{(i)}|Z=z)\le16n^{5/4}/(m\rho)\le16n^{-9/4}
\end{equation}
fails with probability at most $1/8$ under $z\sim Z|B$.
Finally, the assumed conditional intersection bound and Markov's inequality give
\begin{equation}\label{eq:conditional-intersection}
 \Pr_{\mu_z}[\exists j\in\text{epoch }i:S_Q[j]=x_j]\le s/(8n)
\end{equation}
with failure probability at most $1/8$ under the same distribution.
The total failure probability is at most $3/4+o(1)<1$,
so all conclusions hold simultaneously for some $z$ in the support of
$Z|B$.
\end{proof}

\subsubsection{Step 3: a nearly monochromatic rectangle}
Fix $z$ as in \cref{clm:conditioning}.
By \cref{fact:pinsker} and \eqref{eq:near-product},
\[
 \norm{\mu_z-\mu_z^S\otimes\mu_z^X}_{TV}\le3n^{-9/8}.
\]
The product distribution assigns probability at most
$s/(8n)+3n^{-9/8}<s/(4n)$ to intersecting pairs,
by \eqref{eq:conditional-intersection}.
\subsubsection{Step 4: a corruption bound}\label{sec:intersection}

The sets represented by $V,W\in[n]^s$ intersect exactly when
$V_j=W_j$ for some $j$.

\begin{lemma}[Corruption bound]\label[lemma]{lem:intersection}
Let $1\le s\le n$ and let $V,W$ be independent random vectors in $[n]^s$.
If $H(V),H(W)\ge s\log n-s/100$, then
\[
 \Pr[\exists j:V_j=W_j]\ge\frac{4s}{5(2n+s)}\ge\frac{s}{4n}.
\]
No independence between coordinates of either vector is required.
\end{lemma}
\begin{proof}
For prefixes in the support, define the nonnegative conditional Shannon entropy deficits
\[
 a_j(v_{<j})=\log n-H(V_j|V_{<j}=v_{<j}),\qquad
 b_j(w_{<j})=\log n-H(W_j|W_{<j}=w_{<j}).
\]
The entropy chain rule gives
\[
 \sum_j\E_{V_{<j},W_{<j}}\!\left[
   a_j(V_{<j})+b_j(W_{<j})\right]\le s/50,
\]
where the expectation is over the prefixes induced by the distributions of
$V$ and $W$.
Call a prefix pair good at $j$ if $a_j+b_j\le1/10$.
Summing Markov's inequality over $j$ gives
\begin{equation}\label{eq:bad-prefix}
 \sum_j\Pr[\text{prefix pair is bad at }j]\le s/5.
\end{equation}
Conditioned on both prefixes, $V_j$ and $W_j$ are independent,
since the two full vectors are independent. On a good prefix pair,
let their distributions be $\nu,\nu'$. Pinsker's inequality, the triangle inequality and Cauchy--Schwarz give
\[
 \norm{\nu-\nu'}_{TV}\le\sqrt{(a_j+b_j)/\log e}
 \le\sqrt{1/(10\log e)}.
\]
Thus Cauchy--Schwarz gives
\[
 \Pr[V_j=W_j|V_{<j},W_{<j}]
 =\sum_x\nu(x)\nu'(x)
 \ge\frac{(1-\norm{\nu-\nu'}_{TV})^2}{n}\ge\frac1{2n}.
\]
Let $F_j$ be the event that $V_t\ne W_t$ for all $t<j$, and put
$q=\Pr[\exists j:V_j=W_j]$. The event $F_j$ is determined by the two
prefixes and has probability at least $1-q$. Summing over the first
coordinate at which $V$ and $W$ agree and using \eqref{eq:bad-prefix},
\[
 q\ge\frac1{2n}\sum_j\Pr[F_j\text{ and good at }j]
 \ge\frac{s(1-q)-s/5}{2n}.
\]
Rearranging gives the first bound, and $s\le n$ gives the second.
\end{proof}

\begin{proof}[Proof of \cref{lem:robust}]
Steps~2 and~3 give a product distribution whose marginals have entropy
at least $s\log n-s/100$ and which assigns probability less than
$s/(4n)$ to intersecting pairs. This contradicts \cref{lem:intersection}.
\end{proof}

\subsection{Completing the dynamic bounds}\label{sec:completing}

\begin{proof}[Proof of \cref{thm:disj,thm:total}]
Take $n=N^{1/4}$ and $m=N\ge n^4$, and embed
$[n]\times[n]$ into $[N]$. Use the product distribution of
\cref{sec:dist} and keep the random tape $R$ random.
We show that a failure of either bound gives parameters satisfying
\eqref{eq:costs} and \eqref{eq:fast}.

\emph{Update--query trade-off.}
In the single-query setting, the expected costs satisfy \eqref{eq:costs}
with $a=t_u$ and $b=\max\{1,t_q\}$.
Because the number of updates is not given in advance, each query
simulated after a prefix has the same distribution as the query on
that legal input.

If the bound failed, some infinite subsequence would satisfy
\[
 t_q=o\!\left(\left(\frac{\log n}{\log(t_uw)}\right)^2\right).
\]
Because $t_u=n^{o(1)}$,
the expression diverges, so replacing $t_q$ by $\max\{1,t_q\}$
preserves this assumption. The hypotheses \eqref{eq:fast} hold.

\emph{Total work.}
In the interleaved setting, issue an independent uniform query after
each update. This distribution is
fixed independently of the data structure and its operation costs.
If the total-work bound failed, pass to a subsequence with expected
cost at most $n\tau$, where $\tau=o((\log n/\log\log n)^2)$.

Set $a=\max\{1,\tau\}$ and $b=2a$.
The full interleaved cost vector is independent of $T$, and
\[
 \E_{\vS,X,\mathbf Q,R,T}[\lambda_T]
 \le\frac2n\E_{\vS,X,\mathbf Q,R}\!\left[\sum_t c_t\right]
 \le2\tau\le b.
\]
Thus \eqref{eq:costs} holds. Moreover
$\log(aw)=\Theta(\log\log n)$ and \eqref{eq:fast} follows.
The execution retains all earlier query writes and counts their
probes, so the bound counts the entire interleaved execution.

In either case, \cref{thm:simulation,lem:robust} give a contradiction.
Since $\log N=\Theta(\log n)$, this proves both stated bounds.
For the final assertion of \cref{thm:disj}, if $t_u\ge\log^2 N$,
the bound on $\max\{t_u,t_q\}$ is immediate; otherwise
$\log(t_uw)=O(\log\log N)$ and the trade-off proves it.
\end{proof}

\section{Graph applications}\label{sec:applications}

The reductions of \Patrascu~\cite[Appendix~A]{patrascu_mihai_towards_2010}
give the following graph applications; see also \cite{ko_omegalog_2026},
and \cite[Lemma~3.3]{henzinger_unifying_2015} for a similar reduction to
triangle detection.
We recall the constructions and their operation counts.
Throughout, $V$ is the number of vertices and $w=\Theta(\log V)$.
The initial graph and active set are preprocessed for free; all subsequent
probes are counted, with the error convention of \cref{sec:model}.
Only the operations after preprocessing are amortized: if the initial
edges were charged as insertions, incremental reachability to a fixed sink
would have $O(1)$ amortized time by backward search.

\begin{corollary}[Total work for graph problems]\label[corollary]{cor:graph-total}
Each of the following problems has a fixed distribution over its
preprocessing input and a sequence of $\Theta(V^{1/4})$ updates and
$\Theta(V^{1/4})$ queries on which every randomized cell-probe data
structure with the stated error guarantee makes
\[
 \Omega\!\left(V^{1/4}
       \left(\frac{\log V}{\log\log V}\right)^2\right)
\]
probes in expectation:
\begin{enumerate}[label=(\alph*)]
\item \emph{Incremental directed reachability:} insert edges and test
whether a queried vertex can reach a fixed sink. The graph is always
a directed acyclic graph of depth at most two.
\item \emph{Fully dynamic triangle detection:} insert or delete an
edge of a simple undirected graph and test whether the graph contains
a triangle.
\item \emph{Incremental subgraph connectivity:} activate vertices of
a fixed undirected graph and test connectivity between a queried active
vertex and a fixed active vertex in the induced active subgraph.
\end{enumerate}
In particular, if every post-preprocessing sequence with $u$ updates
and $q$ queries has expected total cost at most $ut_u+qt_q$, then
$\max\{t_u,t_q\}=\Omega((\log V/\log\log V)^2)$ for each problem.
\end{corollary}

\begin{proof}
Set $N=(V-1)/2$, $n=N^{1/4}$, and
use the Disjointness distribution of \cref{thm:total} with $m=N$.
Each construction has query vertices $r_1,\ldots,r_N$, element vertices
$v_1,\ldots,v_N$, and a vertex $z$. The incidence edges join $r_q$ to $v_x$ for $x\in S_q$.
The graphs have $O(Nn+N)=O(V^{5/4})$ edges. Each graph query below
tests whether $S_q\cap X\ne\emptyset$, so we complement its answer.

\paragraph{Reachability.}
Orient the incidence edges from query vertices to element vertices. Inserting $x$ adds
$v_x\to z$, and query $q$ tests whether $r_q$ reaches $z$.
The graph remains a depth-two DAG.

\paragraph{Triangle detection.}
Use undirected incidence edges, and insert $\{v_x,z\}$ with each $x$.
For query $q$, insert the edge $\{z,r_q\}$, test for a triangle,
then delete this edge. Without this edge the graph is bipartite;
with it, the triangles are exactly $z,r_q,v_x$ for $x\in S_q\cap X$.

\paragraph{Subgraph connectivity.}
Use the undirected incidence edges and all edges $\{v_x,z\}$.
Initially only the query vertices and $z$ are active; inserting $x$ activates
$v_x$. Query connectivity between $r_q$ and $z$. If $S_q\cap X$ is
nonempty, they share an active neighbor; otherwise $r_q$ is isolated
in the active subgraph.

\paragraph{Error and total cost.}
The reductions map fixed source sequences to fixed target sequences,
independently of answers and random coins, and preserve the error guarantee for each query of every fixed input sequence. Source sequences with distinct insertions, which are all that \cref{thm:total,thm:disj} use, map to legal target sequences. Retain all target memory writes and charge every probe,
including the insertion and deletion of $\{z,r_q\}$ in the triangle reduction; these belong
to the source query, as permitted by \cref{thm:total}. The simulation
adds at most $O(1)$ probes per source operation.

The image of the source distribution is fixed and has $n$ updates and
$n$ queries for reachability and subgraph connectivity, and $3n$ updates
and $n$ queries for triangle detection. Thus \cref{thm:total}, together
with $\log V=\Theta(\log N)$, gives the displayed bound. The assumed
cost guarantee bounds these totals by $n(t_u+t_q)$ or $n(3t_u+t_q)$,
which proves the last assertion.
\end{proof}

\begin{corollary}[Query trade-offs for incremental graph problems]
\label[corollary]{cor:graph-query}
For incremental directed reachability and incremental subgraph
connectivity in \cref{cor:graph-total}(a),(c), suppose a single query
may follow any update prefix, whose length is not given in advance. If the expected cost of the
first $u$ updates is at most $ut_u$ for every fixed input and $u$,
where $1\le t_u=V^{o(1)}$, and the query has expected cost at most $t_q$
and error at most $\delta$ on every fixed input, then
\[
 t_q=\Omega\!\left(
      \left(\frac{\log V}{\log(t_uw)}\right)^2\right).
\]
\end{corollary}
\begin{proof}
For these two reductions, each source operation becomes one target
operation of the same type. The number of target updates is not given in advance, and expected
costs are preserved up to constant probe overhead. Apply
\cref{thm:disj} with $N=\Theta(V)$.
\end{proof}

\section*{AI methodology}
This work was undertaken as part of the author's experiments with AI (ChatGPT-6.0 Astra), motivated by the question of how far AI can go in proving dynamic cell-probe lower bounds with the 2.5-round communication game as a starting point without any human intervention. The author selected the amortized extension of the existing 2.5-round framework as a deliberately accessible test case, expecting it to be within the capabilities of current AI tools. The author's initial prompt for the zero-error proof-development task was:
\begin{quote}
``The goal now is to show $\widetilde{\Omega} ( \log^2 n )$ amortized update bound. \cite{weinstein_amortized_2016} managed to transfer Larsen's bound to amortized bound. The new goal is to transfer \cite{ko_omega_2026}, \cite{ko_omegalog_2026} to amortized bound.''
\end{quote}
The AI developed the initial zero-error amortized lower bounds and their proofs from this prompt. The author subsequently proposed extending the results to bounded error. The bounded-error proofs were developed by the AI and checked through independent AI audits. 

The author takes responsibility for the results presented here and for cleaning up the exposition.

% Use the included bibliography for a self-contained arXiv build.
% To rebuild from references.bib, remove main.bbl and run BibTeX.
\IfFileExists{main.bbl}{%
  % Preformatted from the cited entries in the supplied references.bib.
% Typeset directly for arXiv; this file is not generated by BibTeX.

%
}{%
  \bibliographystyle{alpha}%
  \bibliography{references}%
}

\end{document}